# Content Based Multimedia Information Retrieval To Support Digital Libraries

Mohammad Nabil Almunawar

Faculty of Business, Economics & Policy Studies
Universiti Brunei Darussalam
e-mail: nabil.almunawar@ubd.edu.bn

## Abstract

Content-based multimedia information retrieval is an interesting research area since it allows retrieval based on inherent characteristic of multimedia objects. For example retrieval based on visual characteristics such as colour, shapes or textures of objects in images or retrieval based on spatial relationships among objects in the media (images or video clips). This paper reviews some work done in image and video retrieval and then proposes an integrated model that can handle images and video clips uniformly. Using this model retrieval on images or video clips can be done based on the same framework.

## 1. Introduction

The advancement of information and communication technology creates lots of convenience and opportunities which sometime unthinkable several years ago. The development of the Internet allows people to find information easily in just a matter of clicking a mouse or pressing several characters in the keyboard. Many libraries now offer internet-based search facilities that allow searching documents from home. In fact there are services that provide virtual-library where all documents are converted to digital forms, which definitely can be transferred through computer networks, telephone lines or the Internet.

Currently publishers provide digital versions of their books, journals of magazines, even though most of the publishers provide on-line versions for subscribers only. The on-line services for document searching such as search engines on the Internet or commercial document search engines for finding journals are growing fast and well-known to us now. We are in the digital library era.

Most of the current retrieval facilities provided by library search engines or commercial on-line services are text-based. The text-based searching technology is well developed which can be applied for structure, semi-structure or even unstructured texts. However, information is not only in the form of texts. Everyday we are feed with much more non-text information such as images, video, animation and sound.

Several years ago, multimedia information (integrated of several forms of information including texts, images, video, animation and sound) is a hot topic discussed in many conferences and journals. Until now, multimedia is still a hot topic for research. How can we retrieve multimedia information? A simple approach for retrieving multimedia information is using the well-know text retrieval technology. This means all non-text forms of multimedia information will be given a text annotation/script describing their contents and retrieval or search is based on the text annotations. Although this approach is relatively easy to implement, it has several shortcomings such as tedious work for manual annotations, interpretation problem since different people have different interpretation on the same image for example. A much better approach, which is attracting attention many researchers, is automatic/semi-automatic multimedia content extraction and content-based retrieval. This paper concentrates on the later approach, especially on content-based image retrieval. The organisation of this paper is as follows: in Section 2 we review content-based multimedia retrieval, we discusses object-relationship model in Section 3. In Section 4 we discuss the retrieval system based on the object-relationship model and finally Section 5 concludes the paper.

## 2. Content-Based Multimedia Retrieval

This section discusses content-based multimedia retrieval, especially images and video retrieval. Voice or sound retrieval is an interesting research area. A simple approach is to use the transcript of voice if the sound is news, speech and conversations. Voice recognition is a quite well developed area, so automatic generation of voice transcripts can be done; hence retrieval can be based on the transcripts. Because of space limitation and also there is not much research on voice retrieval, this section will concentrate on images and video retrieval.

### 2.1. Content-based image retrieval

Content-based Image retrieval can be treated in many different ways. One approach is to consider visual properties of images such as colour, shape and texture. This allows us to ask queries such as "Find images that are mainly blue in colour" or "Find all images that have mix colour similar to this example image (pointed by a mouse on the screen)". Also, we can ask the query "Find all images that has similar shape to the shape drawn in the window". The visual feature has been actively pursued.

Another approach is to consider the semantic composition of images in term of the individual objects contained in them and spatial relationships among these objects. This would enable us ask queries such as "Find images the show the red apple *on* the table" or "Find images that show a river running *beneath* a rough stone bridge". Note that these queries incorporate visual properties as well as spatial relationships between objects.

One of the major features of content-based image retrieval is retrieval based on similarity. The retrieval engine uses the concept of similarity retrieval to answer queries; that is the retrieval engine will retrieve candidate answers which are similar to the query. The final decision as to which candidate is appropriate is left to the user in a second stage of filtering. The rest of this sub section reviews briefly research on content-based image retrieval based on similarity concepts.

QBIC (Query By Image Content) [Barb93, Flick95, Nibl93] uses colour, texture and shape to retrieve similar images. Similarity matching based on both average colour and colour histogram is used. For average colour, the similarity measure is based on the weighted Euclidean distance in the colour spectrum. Texture similarity matching is based on weighted Euclidean distance in a three dimensional space (coarseness, contrast, and directionality). Similarity matching of shape is based on the weighted Euclidean distance of several shape features such as circularity, eccentricity, major axis orientation and set algebraic moment invariance. For image sequence retrieval (in video database), the QBIC approach is to generate a representative frame. Once representative frames have been generated, video retrieval is based on the representative frames using the QBIC image retrieval techniques.

Hyperbook on birds [Tabu89, Tabu91] uses the concept of similarity retrieval for retrieving images of birds. It has three query modes: silhouette, birdcall and sketch-of-scene. In silhouette retrieval, the user enters queries interactively by selecting a silhouette from a list of silhouette patterns presented by Hyperbook. The user then selects modifications to the pattern, and the modified pattern is then used as the basis for scanning the database of images. Modifications to the pattern include focus on particular body region, adjust axis of body, adjust size of head, etc. The similarity metric is based on a linear sum of distances from each of the components of the pattern.

Jagadish [Jaga91] proposed shape similarity retrieval based on two-dimensional rectilinear (rectangular covers) shapes representation. In this representation an object (a polygon) is divided into several rectangular shape objects. The similarity measure uses the concept of area-difference. Two shapes are similar if the error area (where the two do not match) is small when one shape is placed on the top of the other. In other works [Gros90, Gros92] the shape of an object is represented by ordered sequence of boundary components, which are feature vectors. Similarity retrieval is based on the minimum cost of converting one feature to the other (error-correction substring matching).

Mehrota [Mehr93, Mehr95] used a set of structural components in the form of maximal curvature boundary points arranged as an ordered set. Shape features are represented as points in a multidimensional space. The similarity of any given two features is defined by the Euclidean distance between the corresponding pair of points in the multidimensional space.

Lu [Lu96] proposed a representation of shape using normalised chain codes [33, 34] of an object's boundary so that the chain codes are invariant under scale and rotation. Using the normalised chain codes, a similarity measurement based on a normalised grid value (1 if a grid is covered by a boundary, 0 otherwise) is proposed.

Scassellati [Scas94] compared several shape similarity measures, including algebraic moments, spline curve distances, cumulative turning angle, sign of curvature and Hausdorff distance. His results are quite interesting. The curve distance (parametric curve distance) is not a good similarity measure for any shape tested. The moment method performs either very well or very poorly on testbed shapes. This method is particularly good for rectangular and arch objects, but it performs very poorly for snake, triangular plane, letter-T and letter-S shapes. The curvature method worked well on shapes that involve protrusion from a central mass. It is not good for shapes that have no protrusion or when there is no central body and it is very bad for T-shapes, squares and ellipses. Hausdorff's distance method performs very well when the test shape is very close to tested shapes but not very good if some flexibility is needed. Scasssellati concluded that the turning angle seems the best one from the above methods since it performs well on any shape tested.

Image retrieval based on colour has been proposed by several researchers [Chua94, Lu94, Meht94, Stri95]. Chua at *al.* [Chua94] have used the colour-pairs technique for fuzzy object-level image retrieval. The basic idea of this technique is to extract a set of distinct colour pairs from an image. This set of colour-pairs is used to model closeness between regions and objects in the image. Similarity of two images can be computed based on sets of colour-pairs, that is if two images contain two similar sets of colour-pairs then the contents of the images are considered similar. Mehtre *et al* [Meht94] have used colour histogram intersections and proposed a distance method and a reference colour method for measuring similarity. In [Mehr93] a quad-tree representation of colour called multi-level colour histograms was proposed. The quad-tree representation can handle positional information on colour, which is its main advantage over ordinary colour histograms. The similarity between two images is computed based on colour histogram intersection.

Similarity retrieval of images encoded in a 2D-string is based on subsequence matching techniques. In other words, the problem of determining the similarity between a query image and a database image is transformed into a 2D-string subsequence matching problem. Chang [Chang87] defines 2D-string matching based on the rank symbols in a 2D-string. Based on this, three types of matching, type-0, type-and type-2, are defined. The matching algorithm is based on subsequence matching. Lee [Lee89] defines similarity matching on 2D-string based on the longest common subsequence in the string. He reduces 2D-string matching to clique finding in an undirected graph. In [Chen93] a modification of Lee's algorithm for similarity match of 2D-string is proposed which claims to perform better than Lee's algorithm.

In [Taka92], Takahashi uses orientation (directional relationships) to retrieve pictures. Since directional relationships such as *right, left* or *west, east* are vague fuzzy representation of spatial relationships using a heuristic function of geometry quantities is proposed. In retrieving images, there is no direct similarity measure between images since an exact matching retrieval method is used. However, since an image is also encoded based on spatial relationships, the spatial relationships that are specified in a query will be matched to spatial relationships in images stored a database based on the estimated degree of fuzziness of the relationships.

Yu [Yu94] used an E-R model to represent an image. The entity part represents the objects in an image together with their attributes and the relationship part represents the relationships between objects in an image. A relationship can be in the form of a spatial relationship as well as a non-spatial relationship called an action relationship.

Gudivada [Gudi95] represents spatial relationships using angles among objects in an image. An image is represented as a graph, called spatial-orientation graph, where nodes are symbol of objects in the image and edges are weighted based the angle between objects The similarity between images are measured based on angles differences among objects in images. The advancing of this technique is that it is invariant to rotation.

We developed powerful unified spatial relationships called 2D-PIR (2-Dimensional Projection Interval Relationship) [Nabi95]. We further used 2D-PIR for image similarity retrieval based on 2D-PIR [Nabi96]. 2D-PIR similarity retrieval relies on object identities supplied by a user during image insertion to an image database. Hence visual properties of objects in an image are ignored. However, the graph-based framework of 2D-PIR can be extended to incorporate visual properties of images such as colour, shape and texture. We will discuss 2D-PIR model and its extension in section 3.

**2.2. Content-based video retrieval**

There are quite many video retrieval systems based on textual annotations describing the content of video clips such as described in [Chua95, Hjel94, Litt91, Oomo94 and Weis95]. But this is not content-based video retrieval. There are several researches focusing on content-based retrieval on video clips. For example,

Zang et al. [Smol94,Tabu91] use visual feature such as colour and shape of representative images for retrieval and browsing video clips in addition to retrieval based textual description of video clips.

Nagasaka and Tanaka [Naga92] used colour differences between frames to grabs cuts or clips in a video. Once a clip has been decided, a representative frame is chosen as an index. Video clip retrieval is based on the colour of specified objects.

Khokhar and Ghafoor [Khok94] used a heterogeneous processing framework to retrieve video clips. The framework consists of image understanding and signal processing algorithms to assist content-based retrieval of video clips. They introduced concept queries such as concept ``slam-dunk'' in basketball. A concept is represented by rules in a knowledge-based. A conceptual query is processed in two steps. Firstly, abstract data representation of video data is used to identify clips which contain intended faces, objects, and audio. Secondly, the selected video clips are analysed at frame level to determine the concept stated in the query.

Retrieving video clips using motion description and trajectories has been investigated in [Dimi94, Dimi95]. The MPEG encoding scheme is used to extract motion vectors of macroblock (a macroblock is 16 x 16 area in MPEG encoding scheme). Based on motion vectors, macroblock trajectories are extracted and then used for object motion recovery. Motion in a video is retrieved using representative trajectories. The trajectory-based retrieval is conducted either exact-match based or similarity-based.

Lee [Lee93] introduced the concept track with multiple class primitives. Video retrieval is conducted on a video-record basis. A video-record is a subsequence of a video sequence in which the starting frame is the frame where one or more objects appearing and the ending frame is the frame where those objects disappear. Each object in a video sequence is described using motion primitives. There are 16 motion primitives which belong to four classes namely translations at constant depth (8 primitives): North(n), North-East(ne), East(e), South-East(se), South(s), South-West(sw), West(w), North-West(nw); translation in depth (2 primitives): Close-the-camera(close) and Away-the-camera(away); rotation at constant depth (2 primitives): Clockwise(cw), Counter-Clockwise(ccw); rotation in depth (4 primitives): Rotate-to-left(left), Rotate-to-Right(right), Rotate-upward(upward), Rotate-downward (downward). Motion primitives in a set are said to be in conflict if the set contains more than one motion primitive belonging to the same class. A motion track is a sequence of one or some combination of non-conflict motion primitives, for example ((se,close),(ne,away),s,n) is a track. A query on video records can be done by a video identifier, by object(s) contained in video records, by track(s) or by their combinations. Lee [69] mentioned video records could be retrieved using track similarity, but proposed no similarity measure.

A model of moving objects which enables queries that refer to future value of dynamic attributes was proposed by Sistla [Sist97]. The model is called Moving Object Spatio-Temporal (MOST). In addition to ordinary (static) attributes, the model is supported by dynamic attributes which are represented by three sub attributes: value, update time and function. The value of a dynamic attribute is determined by time (update time) and the function associated with it. A database state is a mapping that associates a set of objects to its class. Each database state has an associated time stamp. A database history is an infinite sequence of database states. A query is a predicate over the database history. There are 3 types of queries: instantaneous, continuous and persistent query. An instantaneous query is a query evaluated on history starting at the query entered. A continuous query is a sequence of instantaneous queries at certain times and is evaluated continuously until all conditions are satisfied. A persistent query is a sequence of instantaneous queries at a certain time on the infinite history. MOST is equipped with a temporal logic based query language called Future Temporal Logic. The syntax of the language follows the syntax of SQL but is enriched by temporal and spatial predicates.

We [Nabi01] extend modelling images using 2D-PIR for moving objects in video. The model integrates the representation of individual moving objects in a scene with time-varying relationships between them by incorporating both the notions of object tracks and temporal sequence of 2D-PIRs. The model is supported by a set of operations, which form the basis for moving object algebra. This algebra allows one to retrieve scene and information from scenes by specifying both spatial and temporal properties of objects involved. It also provides operations to create new scenes from existing ones.

## 3. Object-Relationship Model

An image is normally composed of several objects. Each object has a name and visual properties as mentioned above and there is a spatial relationship (2D-PIR) between two objects. So an image can be modelled as a graph where nodes are objects and edges are 2D-PIRs between nodes. Each node contains four properties namely object *name* or simply *name*, shape represented by the object boundary, colour represented by average colour of the object and texture of the object.

### 3.1. 2D-PIR symbolic image model

Let us review 2D-PIR symbolic image model before discussing its extension. 2D-PIR is defined as a triple $(\delta,\chi,\psi)$ where $\delta$ is a topological relationship from the set {*dt,to,ct,in,ov,co,eq, cb*}[1]. $\chi$ and $\psi$ are interval relationships {*<, =, m, o, d, s, f, >, mi, oi, di, si, fi*}[2] where $\chi$ represents interval relationship along x-axis and $\psi$ represents interval relationship along y-axis. A detailed description of 2D-PIR can be found in [Nabi95].

Consider Figure 1a and 1b below.

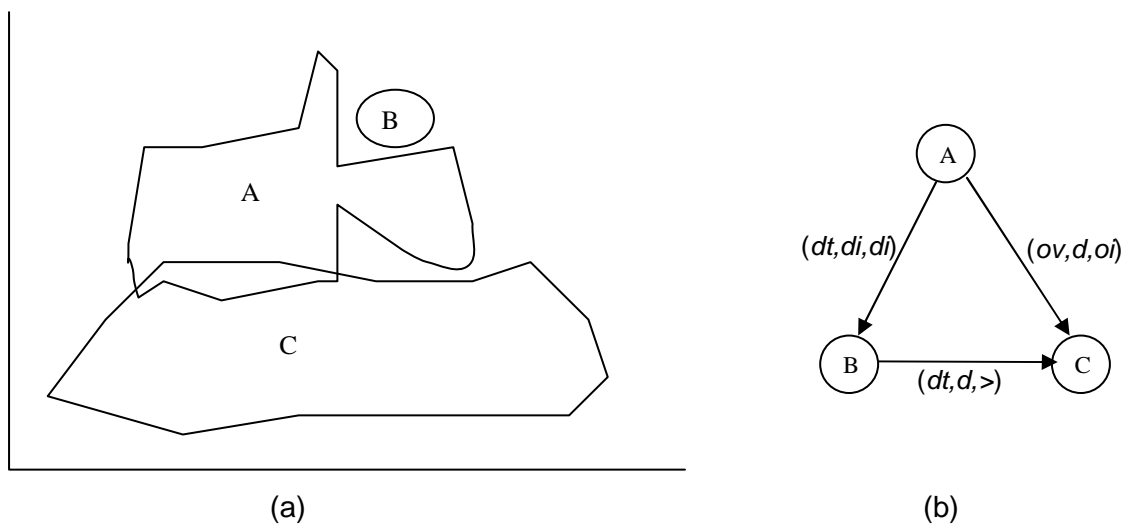

(a)          (b)

---

[1] These represent the topological relationships *disjoint, meets, contains, insides, overlaps, covers, equal,* and, *coveredBy.* See [Egen91] for detail.
[2] These represent the interval relationship *before, equal, meets, overlaps, during, start, finish, after, meet-inverse, overlap-inverse, during-inverse, start-inverse,* and *finish- inverse.* See [Alle83] for detail.

Figure 1. An example of image of three objects (a) and its corresponding 2D-PIR symbolic representation (b).

In Figure 1a the 2D-PIR relationships between object A, object B, object A, object C and object B, object C are (*dt,di,di*), (*ov,d,oi*) and (*dt, d, >*). Figure 1b is the corresponding symbolic image of Figure 1a, which basically a graph (called 2D-PIR graph). 2D-PIR graph (symbolic image model) can be represented as a *bi-list* (2 highly related lists) where the first list contains object symbols or object name and the second list contains the corresponding 2D-PIR relationships among object in the first list. So, the symbolic image of Figure 1b can be represented as ([A,B,C], [((*dt,di,di*), (*ov,d,oi*),(*dt, d, >*)]).

In [Nabi96] we discuss similarity measure on 2D-PIR model and conduct some experiments on similarity retrieval. The result is promising. We developed a prototype image retrieval system based on the above model. Note that the model support reflection and rotation invariant. Table 1 shows the result of experiments on the prototype. Experiments were conducted on a tesbed database containing 400 varieties of images.

Table 1. Recall/Precision Table (P: precision, R: Recall)

| Collection | Accuracy/threshold value | | | | | | | | | | | |
|---|---|---|---|---|---|---|---|---|---|---|---|---|
| | 0 | | 20 | | 40 | | 50 | | 60 | | 80 | |
| | R | P | R | P | R | P | R | P | R | P | R | P |
| Sydney scenery | 100 | 29 | 100 | 38 | 94 | 78 | 66 | 89 | 28 | 98 | 8 | 100 |
| Antelopes | 87 | 32 | 83 | 41 | 67 | 74 | 60 | 82 | 35 | 100 | 21 | 100 |
| Falcons | 90 | 52 | 90 | 72 | 79 | 78 | 79 | 78 | 69 | 82 | 50 | 100 |
| Sunset | 89 | 25 | 77 | 31 | 53 | 74 | 48 | 77 | 29 | 94 | 20 | 100 |
| Monkeys | 100 | 20 | 100 | 29 | 50 | 100 | 50 | 100 | 25 | 100 | 25 | 100 |
| Hawks | 95 | 50 | 95 | 53 | 79 | 91 | 69 | 96 | 43 | 100 | 24 | 100 |
| Owls | 96 | 33 | 84 | 45 | 81 | 90 | 74 | 94 | 62 | 100 | 24 | 100 |
| Mountains | 100 | 14 | 100 | 26 | 74 | 56 | 51 | 76 | 30 | 100 | 23 | 100 |
| Geometric Mean | 95 | 32 | 91 | 41 | 72 | 80 | 62 | 87 | 40 | 97 | 24 | 100 |

There are different accuracy of 2D-PIR matching values or threshold values in Table 1: 0, 20, 40, 50, 60 and 80. Zero means matching is solely based on symbols or objects' names; in other words, it is keyword-based matching. We use zero to compare similarity matching using 2D-PIR with keyword-based matching. The accuracy 20, 40, 50, 60, and 80 thresholds are use to show the effect of 2D-PIR on recall and precision of similarity retrieval (note that a user can adjust the threshold value according to his/her preference).

It can be seen (from Table 1) that on average (geometric mean) the recall value is 95 percent for keyword matching. However, at the same time, the precision is low. When the threshold value is increased, the recall value is decreased, but the precision value increases. For example, the precision value increase by factors 2.7 and 3.0 for threshold values of 50 and 60 respectively.

**3.2. Examples similarity retrieval using 2D-PIR**

Figure 2 is the result of query on a collection of sunset pictures using similarity threshold 30. Note the top left is the query. Note also they are ordered by decreasing similarity, the top left being 100% by definition. Figure 3 shows the result of query on the same collection using similarity threshold 60. The result shows that the system retrieve similar images especially when the threshold value is quite high. Figure 4 and 5 are sketch query and the corresponding similarity result. Finally Figure 6 is the test to show the 2D-PIR model support rotation and reflection.

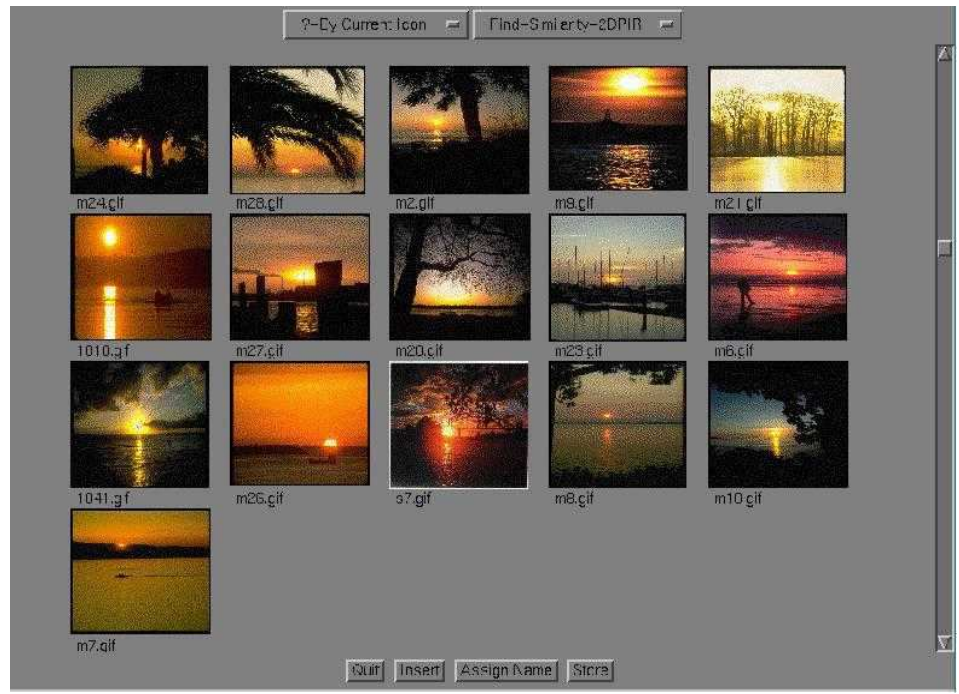

Figure 2: An example similarity retrieval on sunset images (to left image in the query); threshold value: 30.

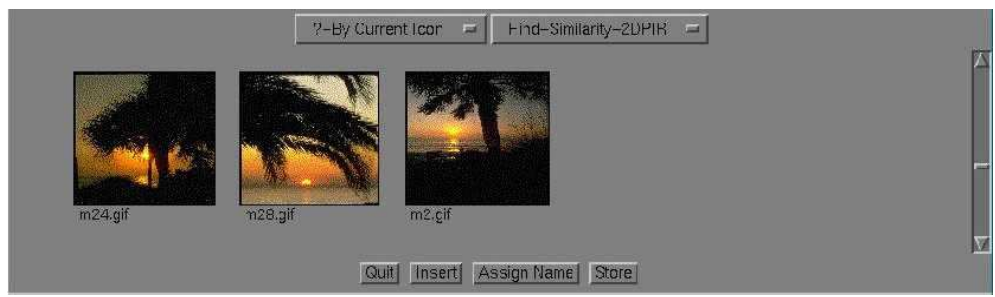

Figure 3: Similarity retrieval result on sunset images (to left image in the query); threshold value: 60.

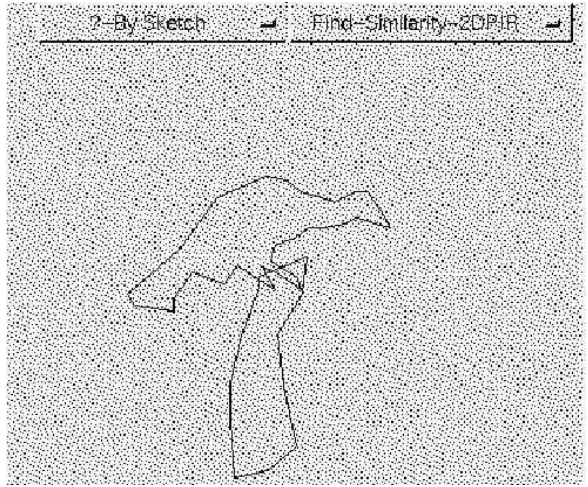

Figure 4: An example of sketch query: find a hawk perching on a stump

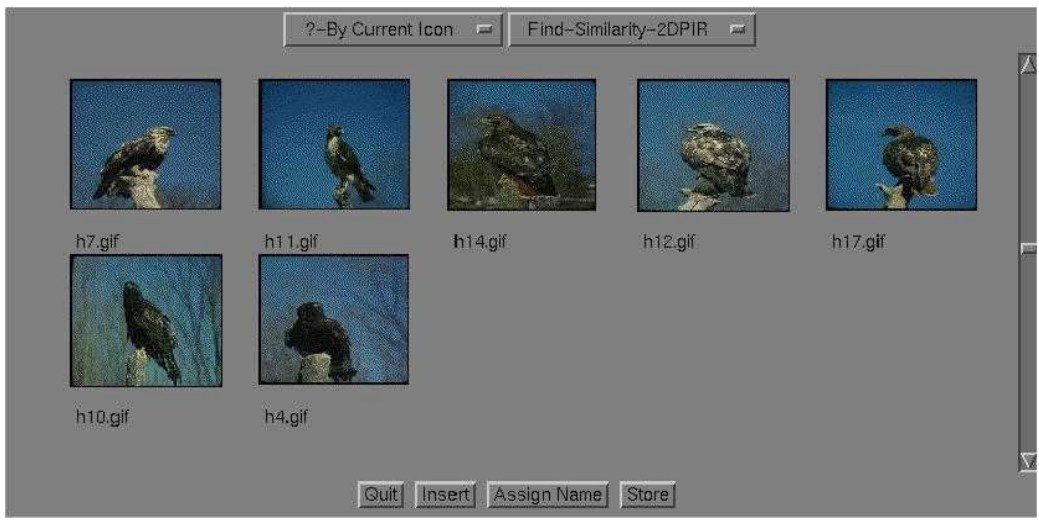

Figure 5: Result of the sketch query of Figure 5.

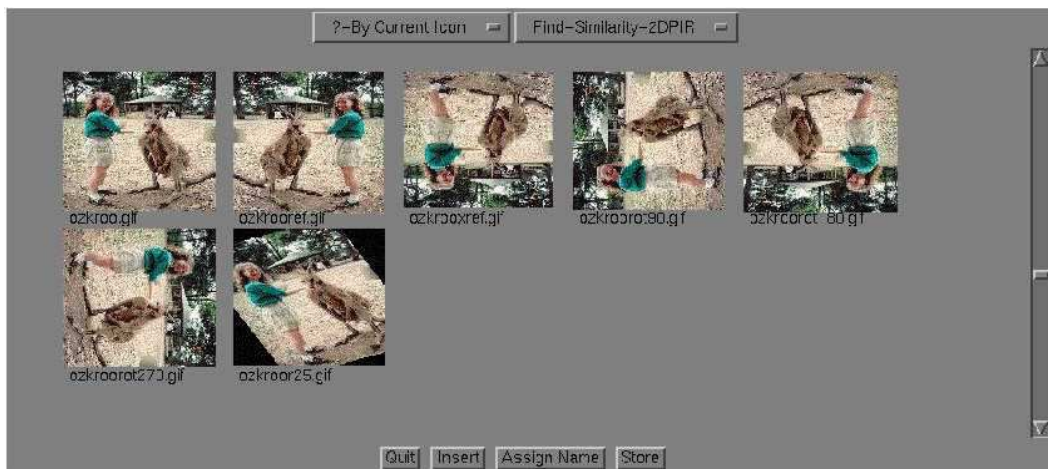

Figure 5: A test query to show 2D-PIR supports reflection and rotation images

### 3.3. A extension of 2D-PIR

Ideally, an image retrieval system provides many ways to retrieve intended images. 2D-PIR model provide keyword based and integrated keyword-spatial relationships (2D-PIRs) similarity retrieval.

Several image similarity retrievals (see section 2) use visual feature such as colour, shape and texture. There is a prominent shortcoming similarity retrieval solely based on visual feature which is not mentioned in the literature, either they work for whole image ignoring objects in the image such as retrieval based on colour or texture or it only considers one object in an image (retrieval based on shape).

Our 2D-PIR image model offers extensible modelling of image. Visual feature mention above can be incorporated in the model easily. We believe that if visual feature incorporated in the model then the model will offer flexible yet powerful similarity retrieval on images. We will reveal the extension model shortly, but it is beyond the scope of this paper to discuss integrated similarity measure of spatial relationships and visual features.

Recall that 2D-PIR symbolic image can be represented as a *bi-list* where the first list contains object symbols or object name and the second list the corresponding 2D-PIR relationships among object in the first list. To incorporate visual features simple extend the first list to include those features.

Let X be an object name and $X_C$, $X_S$, $X_T$ be its corresponding colour, shape and texture. Note that X, $X_C$, $X_S$, $X_T$ are object attributes. Consider an image contains two objects named as A and B. The *bi-list* representation of this image is $([(A,A_C,A_S,A_T),(B,B_C,B_S,B_T)], [(\delta_{AB},\chi_{AB},\psi_{AB})])$. We call this representation as Object-Relationship Model for an image.

Using Object-Relationship Model, the image representation of Figure 1 is $([(A,A_C,A_S,A_T),(B,B_C,B_S,B_T),(C,C_C,C_S,C_T)],[(dt,di,di),(ov,d,oi),(dt,d,>)])$. Retrieval can be based on object attributes whether it's name or its visual attributes, spatial relationships on the whole representation. This will offer flexible retrieval for users.

### 3.4. An extension of Object-Relationship Model to incorporate moving objects

As we mentioned earlier, our 2D-PIR image model is extensible to moving objects. The detail discussion of moving objects including its algebra based on dynamic 2D-PIR called TS-2D-PIR can be seen in our recent published paper [Nabi01].

We would like to use *bi-list* representation to model moving objects with their visual features. A moving object has its track, which will be represent as list $[(d_i,r_i,I_i)]$ where $d_i$ and $r_i$ are distance travelled and direction during time interval $I_i$ for i = 1 to n. Note an empty list implies the object does not move. The TS-2D-PIR between two objects is represented as list $[(\delta_i,\chi_i,\psi_i, I_i)]$ where $\delta_i, \chi_i, \psi_i$ are 2D-PIRs during the time interval , $I_i$. Note that if the list a singleton list then both objects do not move.

Let A and B two moving objects in a scene or clip. The *bi-list* representation of those moving objects is: $([(A,A_C,A_S,A_T, [(d_{Ai},r_{Ai},I_i)]),(B,B_C,B_S,B_T, [(d_{Bi},r_{Bi},I_i)])], [[(\delta_{AB},\chi_{AB},\psi_{AB}, I_i)]])$.

## 4. The Retrieval System

The architecture of the system is shown in Figure 6. Note that this is the image retrieval system. Moving objects or video clips retrieval system is relatively similar to the image retrieval system. The system composed of five major components: user interface, image transformer, insertion engine, retrieval engine and image database.

### 4.1. User interface

Unlike traditional databases, queries in an image or multimedia database are most likely posed in non-textual forms, such as sketches or existing image (iconic or "thumbnail" versions) or started by a text query followed by non-textual queries. Therefore a Graphical User Interface (GUI) must be supported so that non-textual queries can be posed easily. The Object-Relationship Model for images directly supports graphical querying. Using this scheme, a sketch query is mapped to a *bi-list* representation, which is ready to be compared with pictures stored in a database.

### 4.2 Image transformer

The symbolic image transformer converts an image to the corresponding Object-Relationship Model representation (*bi-list*). An image is considered as a set of polygons representing the objects in the image along with symbols to identify each object and their visual attributes. The image transformer accepts the set of polygons and their corresponding symbols and then grabs objects' shapes and their colour as well as computes 2D-PIR relationships among objects. The symbolic image transformer also accepts a user query sketch as a set of polygons and the corresponding symbols. The output of this transformer is a *bi-list* that represents an image.

### 4.3. Insertion engine

Once the user chooses an image to be inserted into the database, the system will display the image in the main window. The user is required to identify each object in the image by manually drawing their boundaries and assigning a text-based identifier for the object. If all of the objects have been identified, the user instructs the system to store the image. The system (through the image transformer) grabs shapes and average colour of every object and computes the 2D-PIR relationships among the objects then generates a *bi-list* representation for the image and stores this representation along with pointers to the original image and its iconic version.

### 4.4. Search engine

Image retrieval is either *sketch-based* or *image-based.* Sketch-based queries begin by the user drawing a sketch containing rough object-outlines in some spatial arrangement and identifying each object. Image-based queries are commenced by clicking on an iconic image, either from a previous query or from a set of images chosen randomly from the entire image database. In both cases, the query is transformed into a *bi-list,* which is then used as the basis of the matching process.

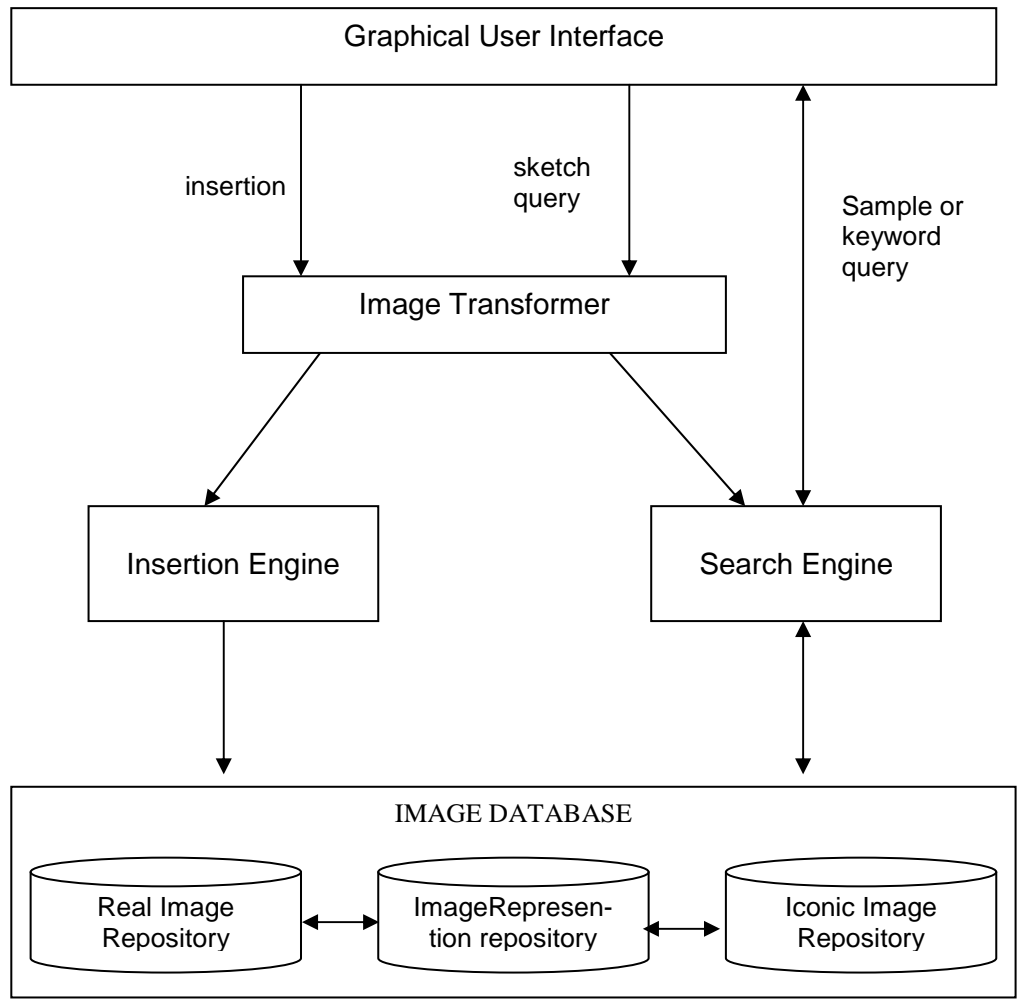

Figure 6: Architecture of a Prototype Image Retrieval System

In our prototype, matching *is approximate.* Approximate (similarity) matching uses the set of objects that are common to both the query and the image and uses the previously defined similarity measure on the spatial relationships between the common objects. In the case of similarity retrieval, the user may adjust the retrieval accuracy, which effectively sets a threshold value on similarity measure. An accuracy threshold of 0 means that any images having at least some of the objects in the query will be retrieved, regardless of the spatial arrangement of the objects. An accuracy of 100 means that only images for which the set of common objects has exactly the same spatial arrangement in both the image and the query will be retrieved. Note that we have not incorporated visual attributes of image in the current prototype.

**4.5. Image database**

Conceptually, the database is partitioned into three parts: a real image repository (where the original full-sized images are stored), an iconic image repository where an iconic version ("thumbnail") of each image is kept) and the image representation repository. Physically, we store only a pointer (URL) to the original image, so, in fact, the system can form the basis of a networked-wide image database. Iconic images are stored locally since they are presented to the user in the first stage of the retrieval process.

The image representations are also stored locally, and are central to the image retrieval: process. Each image representation contains:

- pointers to the original and iconic images

- *bi-list* representation of the image

## 5. Conclusion and Challenges

We are living in the age of information. Obviously we are not lack of information; in fact, we are overloaded by information. Information retrieval (IR) helps us to filter information so that only information with high degree relevancy and accuracy will be retrieved. There are many implementation of IR such as implemented in libraries and Internet search engines.

Most of implementations of IR available are based on text retrieval. However, information is not solely on texts, in fact, daily we are feed by multimedia information through our senses much more than information we get from reading (texts). The volume of multimedia information is growing fast and if they are not organise properly than we cannot benefit from the usefulness of multimedia information that we have.

Retrieval of multimedia information based on textual description of multimedia objects (using text-based IR technology) requires tedious effort in providing textual annotations and provides incomplete and bias information about the content of multimedia objects. Hence an objective content-based or combination of text-based and objective content-based multimedia information retrieval is needed.

We describe our model for multimedia retrieval. The model is powerful and extensible. Our experiments on the original model (without visual features) on retrieval of images showed the model is promising. We believe the extension of the model to incorporate visual features will be much better in term of flexibility and accuracy and relevancy of retrieval.

When we come to perform similarity retrieval from millions of pictures stored in multimedia databases (possibly distributed in the Internet), efficient and flexible modelling and similarity matching are not enough. We also require the overall similarity retrieval process to be efficient. Traditional text-based retrieval uses indexing technique (such as inverted files) to avoid linear search for retrieval. Indexing on multimedia data is much more sophisticated that indexing on texts. There are some indexing researches for individual attribute, but not for integrated features such as proposed in this paper.